\documentclass[]{pasj02} 
\usepackage[switch,mathlines]{lineno}

\jyear{2025}
\Received{}
\Accepted{}
\definecolor{HHcolor1}{rgb}{0.0,0.0,0.0}
\definecolor{HHcolor2}{rgb}{0.0,0.0,0.0}
\definecolor{HHcolor3}{rgb}{0.0,0.0,0.0}

\newcommand{\hhBlue}[1]{{\color{HHcolor3}#1}}

\begin{document} 

\title{Clumpy Outflows from Super-Eddington Accreting Black Holes I: Radiation Hydrodynamics Simulations and Observational Implications}

\author{
Haojie \textsc{Hu},\altaffilmark{1}\altemailmark
\orcid{0000-0003-3143-3995} \email{huhaojie@ccs.tsukuba.ac.jp; (JSPS Fellow)} 
Yuta \textsc{Asahina},\altaffilmark{1,2}\orcid{0000-0003-3640-1749}
Shogo \textsc{Yoshioka},\altaffilmark{1,3} \orcid{0009-0005-6822-4179}
Hiroyuki R. \textsc{Takahashi}\altaffilmark{2}\orcid{0000-0003-0114-5378}
 and 
Ken \textsc{Ohsuga}\altaffilmark{1}\orcid{0000-0002-2309-3639} 
}
\altaffiltext{1}{Center for Computational Science, University of Tsukuba, 1-1-1 Tennodai, Tsukuba, Ibaraki 305-8577, Japan}
\altaffiltext{2}{Department of Natural Science, Faculty of Arts and Science, Komazawa University, 1-23-1 Komazawa, Setagaya-Ku, Tokyo 154-8525, Japan}
\altaffiltext{3}{Department of Astronomy, Graduate School of Science, Kyoto University, Kitashirakawa-Oiwake-cho, Sakyo-ku, Kyoto 606-8502, Japan}




\maketitle

\begin{abstract}
Recent advances in X-ray spectroscopic observation have enabled researchers to reveal distinct clumpy structures in the super-Eddington outflows from the supermassive black hole in PDS 456 \citep{Xrism+2025}, initiating detailed investigation of fine-scale structures in accretion-driven outflows. 
In this study, we conduct high-resolution, two-dimensional radiation-hydrodynamics simulations with time-varying and anisotropic initial and boundary conditions to reproduce outflows launched from super-Eddington accretion flows and analyze their statistical properties.
The resulting clumpy outflows extend across a wide range of radial distances and polar angles, exhibiting typical properties such as a size of $\sim10~r_{\rm g}$ (where $r_{\rm g}$ is the gravitational radius), a velocity of $\sim0.05$–$0.2~c$ (where $c$ is the speed of light), and about five clumps along the line of sight. 
Although the velocities are slightly smaller, these characteristics reasonably resemble those obtained from the XRISM observation.
The gas density of the clumps is on the order of $10^{-12}$–$10^{-13}~{\rm g~cm^{-3}}$, and their optical depth for electron scattering is approximately $1$–$10$.
The clumpy winds accelerated by radiation force are considered to originate from the region within $\lesssim 300~r_{\rm g}$.
\end{abstract}


\section{Introduction}

Accretion onto black holes (BHs) represents the most efficient energy conversion mechanism in astrophysical environments \citep{Shakura+1973}, powering diverse luminous phenomena including X-ray binaries, active galactic nuclei (AGNs), and quasars \citep{Begelman+1984, Rees1984, Kaaret+2017}. During the accretion process, enormous quantities of gravitational potential energy are liberated in the form of radiation, thermal, and kinetic energy \citep{Shakura+1973, Blandford+1977, Fabian2012}. While radiative energy propagates as electromagnetic (EM) waves, mass outflows—comprising high-velocity jets and winds spanning a range of speeds—transport thermal and kinetic energy outward, substantially influencing their surrounding environments \citep{King+2015, Blandford+2019}. Observational evidence indicates that outflows constitute a fundamental component of accretion systems across all scales, occurring regardless of accretion rate or central object mass \citep{Proga+2000, Tombesi+2010, King+2012}.

Among the diverse range of accretion regimes, super-Eddington accretion stands out as a phenomenon of exceptional interest and astrophysical significance \citep{Abramowicz+1988}. This extreme accretion mode has been successfully invoked to explain several important astrophysical phenomena: the rapid growth of seed BHs in the early universe, the extraordinary luminosities of ultra-luminous X-ray sources (ULXs), and the existence of highly luminous quasars \citep{Volonteri+Rees2005, King2009, Inayoshi+2016, Kaaret+2017, Inayoshi+2020}. From a theoretical perspective, super-Eddington accretion arises when a compact object becomes embedded within a dense gas environment where the emergent radiation is effectively trapped in the immediate vicinity of the accretor \citep{Begelman1978, Begelman1979, Ohsuga+2005, Ohsuga+Mineshige2011, Jiang+2014, Inayoshi+2016, Fan+2023}. Under these conditions, radiative energy transport becomes highly inefficient, and outflows emerge as one of the primary channels, alongside radiation, for evacuating thermal and kinetic energy from the system. Consequently, powerful outflows are an inevitable product of super-Eddington accretion processes \citep{Takeuchi+2013, Jiang+2014, Yuan+2014, Piran+2015, Kobayashi+2018, Kitaki+2018, Kitaki+2021, Yoshioka+2022, Hu+2022}. The kinematic properties of these outflows constitute a critical component for interpreting both the structural characteristics of super-Eddington accretion disks and their consequent feedback effects on galactic environments \citep{King+2015, Hu+2022}.

From an observational perspective, AGNs and quasars serve as exceptional astrophysical laboratories for investigating BH accretion processes and their associated feedback mechanisms \citep{Fabian2012, King+2015}. Of particular significance is the subset of luminous quasars observed to radiate above their theoretical Eddington limits, providing valuable opportunities to examine super-Eddington accretion regimes and their resultant outflow phenomena \citep{Du+2014, Shen+Ho2014, Wu+2015, Tombesi+2015, Bischetti+2017}. A compelling example is the radio-quiet quasar PDS 456, situated in the relatively nearby universe at redshift $z=0.184$, which harbors a supermassive black hole (SMBH) of mass $M_{\rm BH}\simeq 5\times 10^8~M_\odot$ accreting at or beyond its Eddington threshold \citep{Amorim+2023}. High-resolution X-ray spectroscopic observations of this system have revealed the presence of remarkably energetic outflows with velocities approaching $\simeq 0.3 ~ c$ \citep{Reeves+2003, Behar+2010, Nardini+2015, Boissay-Malaquin+2019}. These characteristics position PDS 456 as an exemplary target for examining powerful outflow dynamics and their regulatory influence on the co-evolutionary relationship between SMBHs and their host galaxies \citep{Heckman+Best2014}.

The recent advent of high energy spectrum resolution X-ray observations enabled by the XRISM mission has revolutionized our capacity to investigate the fine-scale structure of astrophysical outflows. Groundbreaking observations by \citet{Xrism+2025} have revealed multiple distinct absorption features in the iron K-shell energy band of quasar PDS 456. These spectral signatures are interpreted as arising from absorption by discrete clumps of gas traveling at a range of relativistic velocities ($\simeq 0.2-0.3~c$) \citep{Xrism+2025, Xu+2025}. This evidence strongly suggests that outflows from super-Eddington accreting BHs possess a complex, heterogeneous structure—a significant difference from the conventional single-velocity outflow paradigm that has dominated the field. While theoretical work has previously indicated that clumpy outflows can naturally arise in super-Eddington accretion simulations through thermal and/or radiative instability mechanisms \citep{Takeuchi+2013, Kobayashi+2018}, these new observational constraints motivate crucial questions regarding their formation mechanism, subsequent evolution, and cosmological significance in regulating the co-evolutionary relationship between SMBHs and their host galaxies.

In this paper, we conduct two-dimensional radiation-hydrodynamics (RHD) simulations to examine the generation and subsequent evolution of clumpy outflows from super-Eddington accretion disks around SMBHs. This investigation focuses specifically on modeling the observed clumpy structures within these outflows and characterizing their statistical properties through comprehensive numerical simulations. The manuscript is structured as follows: In Section~\ref{sec:setup}, we detail our simulation methodology, including the governing equations and numerical configurations. Section~\ref{sec:CO} presents the principal findings from our fiducial simulation regarding clumpy outflow characteristics. In Section~\ref{sec:discussion}, we analyze the sensitivity of our results to various numerical parameters and physical assumptions. Finally, in Section~\ref{sec:conclusion}, we summarize our findings and discuss future perspectives.

\section{Numerical Simulations}\label{sec:setup}

\subsection{Basic equations in UWABAMI+INAZUMA code}
 
In this paper, we perform two-dimensional RHD simulations to investigate the formation and evolution of clumpy outflows from super-Eddington accreting SMBHs, utilizing UWABAMI+INAZUMA code \citep{Takahashi+2016, Asahina+2020, Asahina+2024}.
Although this code is a General-Relativistic Radiation-Magnetohydrodynamics (GR-RMHD) code, GR effects do not appear in the present study. In addition, we set the magnetic fields to zero in our simulations, since it has been reported that magnetic fields do not play an essential role in the formation and evolution of clumpy outflows \citep{Takeuchi+2013}. A detailed investigation of their effects is left for future work.

In the simulation, the speed of light ($c$) is taken as unity, \hhBlue{but we restore c in the discussion for physical completeness.} 
Conventionally, Latin scripts indicate space components while Greek subscripts indicate spacetime components. 
The following conservation equations are numerically solved in the simulation:
\begin{equation}
    \frac{\partial (\rho u^t)}{\partial t} + \frac{1}{\sqrt{-g}}\frac{(\sqrt{-g}\rho u^i)}{\partial x^i} = 0,
\end{equation}
\begin{equation}
    \frac{\partial T^t_\nu}{\partial t} +  \frac{1}{\sqrt{-g}}\frac{(\sqrt{-g}T^t_\nu)}{\partial x^i} - T^\kappa_\lambda \Gamma^\lambda_{\nu\kappa} = G_\nu,
\end{equation}
\begin{equation}
    \frac{\partial R^t_\nu}{\partial t} +  \frac{1}{\sqrt{-g}}\frac{(\sqrt{-g}R^i_\nu)}{\partial x^i} - R^\kappa_\lambda \Gamma^{\lambda}_{\nu \kappa}= -G_\nu,
\end{equation}
where $\rho$, $R^\mu_\nu$, $u^\mu$ and $g$ are the gas density, the energy-momentum tensor for radiation, the four velocity, and the determinant of the metric tensor $g_{\mu \nu}$, respectively. As mentioned above, we do not include magnetic field in the simulation. Thus, the covariant magnetic field $b^\mu=0$. The energy-momentum tensor is
\begin{equation}
    T^\mu_\nu = \left ( \rho + \frac{\Gamma}{\Gamma -1}p\right )u^\mu u_\nu + p\delta^\mu_\nu,
\end{equation}
where $p$ is the gas pressure, $\delta^\mu_\nu$ is the Kronecker delta, and $\Gamma=5/3$ is the specific heat ratio, and $\Gamma^\lambda_{\nu \kappa}$ is the Christoffel symbol of the second kind.
$G_\nu$ is the radiation four-force:
\begin{equation}
\begin{split}
    G_\nu = &-\rho \kappa_{\rm abs} (R_{\nu \alpha} u^{\alpha} + 4 \pi B u_\nu) \\
    &-\rho \kappa_{\rm sca} (R_{\nu \alpha} u^{\alpha} + R_{\alpha \beta} u^{\alpha}u^{\beta}u_{\nu}),
\end{split}
\end{equation}
where $\kappa_{\rm abs}=6.4\times 10^{22}\rho T_{\rm gas}^{-7/2}~{\rm cm^2~g^{-1}}$ and $\kappa_{\rm sca} = 0.4 ~{\rm cm^2~g^{-1}}$ are opacities for absorption and scattering in the co-moving frame (see \cite{Rybicki+Lightman2024}). The blackbody intensity $B$ is given by the gas temperature $T_{\rm gas}$:
\begin{equation}
    B = \frac{a_{\rm rad} T_{\rm gas}^4}{4\pi},
\end{equation}
with $a_{\rm rad}$ the radiation constant. The gas temperature is obtained from the equation of state
\begin{equation}
    p = \frac{\rho k_{\rm B}T_{\rm gas}}{\mu m_{\rm p}},
\end{equation}
where $k_{\rm B}$ is the Boltzmann constant, $m_{\rm p}$ is the proton mass, and $\mu=0.5$ is the mean molecular weight.
We employed an M-1 formalism \citep{Gonzalez+2007} to close equation (3). 
The radiation energy momentum tensor is given by 
\begin{equation}
  R^{\mu\nu} = 4p_\mathrm{rad}u^{\mu}_\mathrm{rad}u^{\nu}_\mathrm{rad}+p_\mathrm{rad}g^{\mu\nu}, 
\end{equation}
where $p_\mathrm{rad}$ and $u^{\mu}_\mathrm{rad}$ are the radiation pressure and radiation frame's four-velocity, respectively.

\subsection{Simulation setup \& Initial conditions}
\label{sec:BC}

In the simulation, we use spherical coordinates $(r,~\theta,~\phi)$ for a static, non-spinning BH
with mass $M_{\rm BH}=10^7~M_\odot$. We set inner radius of the simulation box to $30~r_{\rm g}$ with $r_{\rm g}=M_{\rm BH}$ the gravitational radius (the Gravity constant $G$ is set to unity in simulation, \hhBlue{but kept in our discussion}), and outer radius to $1500 ~r_{\rm g}$, covering the observational scales of clumpy winds (see \cite{Xrism+2025}). The simulation box covers polar angles within $5^\circ \leq \theta \leq 90^\circ$, to avoid the occurrence of high-speed jets near the polar axis. 
The number of grid points is set to be ($N_{r}$, $N_{\theta}$) = $(640,~315)$ in space. The grids are uniformly spaced in both $r-$ and $\theta-$directions.

\begin{figure}
 \begin{center}
  \includegraphics[width=80mm]{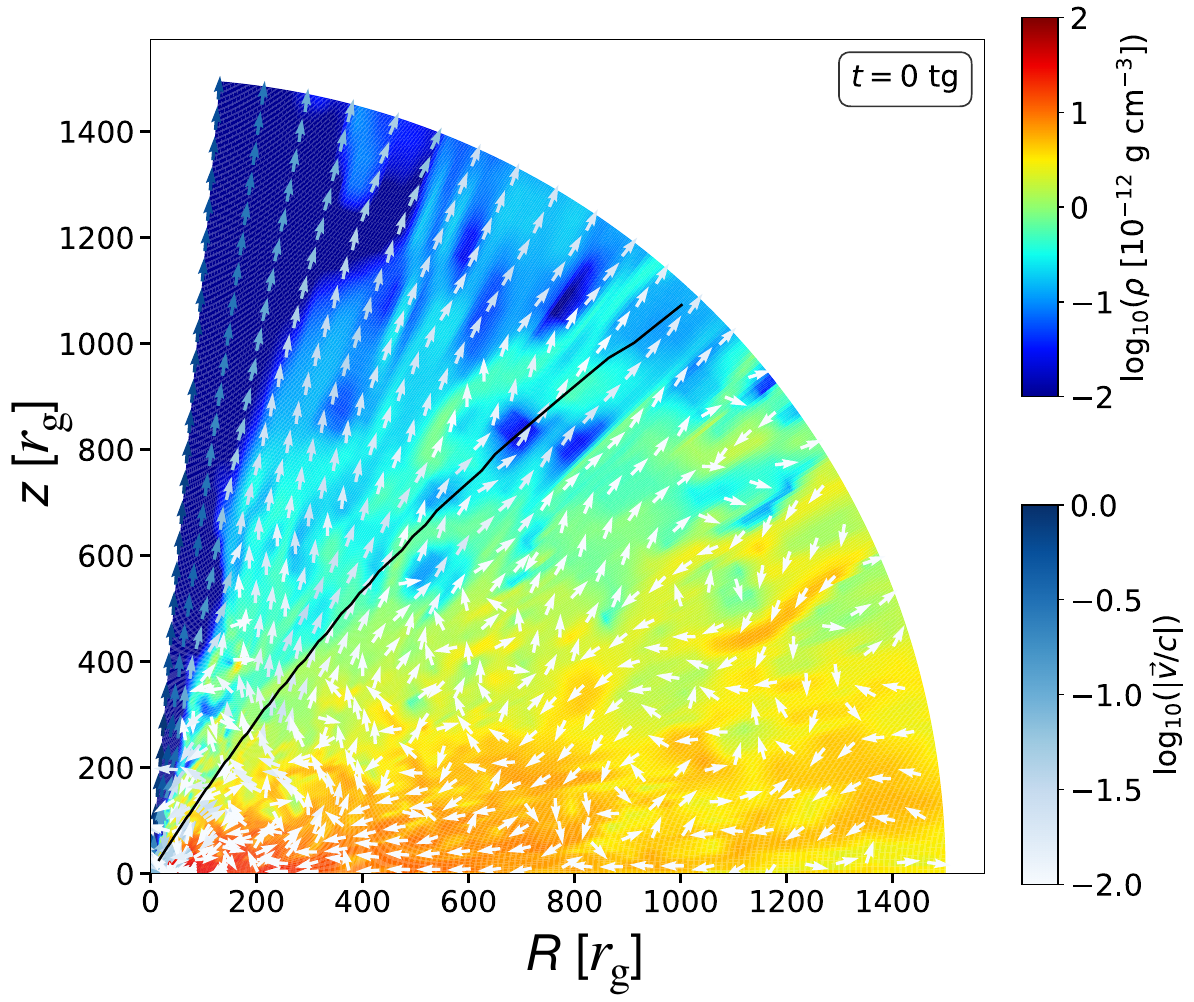}
 \end{center}
 \caption{Visualization of the initial conditions and disk boundary. The 2D density distribution is displayed using pseudo-colors, while velocity vectors are overlaid as arrows. The black solid curve indicates the disk boundary. The cylindrical coordinates are defined as $R=r\cos\theta$ and $z=r\sin\theta$.
 Alt text: Initial Conditions.}
 \label{fig_ICBC}
\end{figure}

For initial conditions, one snapshot data at time $t=4.9\times 10^8~ {\rm s}$ from RHD simulations in \citet{Yoshioka+2024} is adopted. Subsequently, the RHD data are converted into the GR-RMHD data input format and are bilinearly interpolated to match the grid if necessary. 
Here we note that the grid cell size in the present simulations is sufficiently smaller than that in \citet{Yoshioka+2024}, which is advantageous for resolving the detailed structure of the outflows.
The initial density, velocities, and the disk boundary are visualized in Figure~\ref{fig_ICBC}. At this time snapshot, the system has reached a steady state with a mass accretion rate of $\sim 500~L_{\rm Edd}$, where $L_{\rm Edd}$ denotes the Eddington luminosity.
The accretion structure consists of a turbulent dense disk, outflowing winds in the off-disk region (polar angle $\sim 10^\circ$ to the disk surface), and a high-velocity jet near the polar axis ($\theta\leq 10^\circ$).

\subsection{Boundary conditions}
\label{sec:BC}

In the $\theta-$direction, zero-gradient (free) boundary conditions for all physical quantities are adopted at $\theta=5^\circ$ and $\theta=90^\circ$. Meanwhile, to focus specifically on the outflow dynamics rather than the accretion disk evolution, a disk boundary is set in our simulations. 

In simulation, the physical quantities on and below this boundary are taken from the pre-existing simulation data of \citet{Yoshioka+2024}. Here, the boundary of the disk surface is simply given by
\begin{equation}
    \theta=\theta_0 + \frac{1}{2}\cdot (\theta_1-\theta_0)\cdot \left (1+\tanh[{{(r-{\delta r})}}/{\delta r}] \right ),
\end{equation}
with $\theta_0=\pi/12$ and $\theta_1=1.4$ and $\delta r = 2000~r_{\rm g}$.
The disk boundary approximately traces the disk surface defined by \citet{Yoshioka+2024}, where the radiation force and gravity are in balance. In practice, physical quantities are computed using bilinear interpolation to match grid cells in two simulations and between two temporally adjacent datasets. The disk data of \citet{Yoshioka+2024} are updated at a time interval of $\delta t=10~r_{\rm g}/c$.

In the radial direction, outflow boundary conditions are adopted at the outer boundary ($r=1500~r_{\rm g}$), while at the inner boundary($r = 30~r_{\rm g}$), the physical quantities evolve according to the simulation data of \citet{Yoshioka+2024}, in the same manner as for the disk boundary. In a words, the effective computational domain is defined as the region between the inner and outer boundaries, located above the disk boundary. The cases with the inner boundary set at $r = 100~r_{\rm g}$ and $r = 300~r_{\rm g}$ are discussed in Section~\ref{sec:discussion}.

\section{Clumpy outflows: properties and observables}
\label{sec:CO}

\subsection{Overall Structures}\label{sec:3_1}

Following the numerical framework outlined in Section~\ref{sec:setup}, we evolve our simulation for a total duration of approximately $2\times 10^5~r_{\rm g}/c$. This extended temporal domain encompasses more than three dynamical timescales, where the dynamical time is defined as $t_{\rm dyn}\sim r_{\rm out}/v_{\rm Kep}\sim 6\times 10^4~r_{\rm g}/c$ at the outer boundary radius $r_{\rm out}$, with $v_{\rm Kep}$ representing the local Keplerian orbital velocity. The  
system attains a quasi-steady configuration after approximately one dynamical timescale, at which point the initial transient outflows have propagated beyond the computational domain
and the statistical properties become time-invariant.

\begin{figure*}[htbp]
 \begin{center}
  \includegraphics[width=170mm]{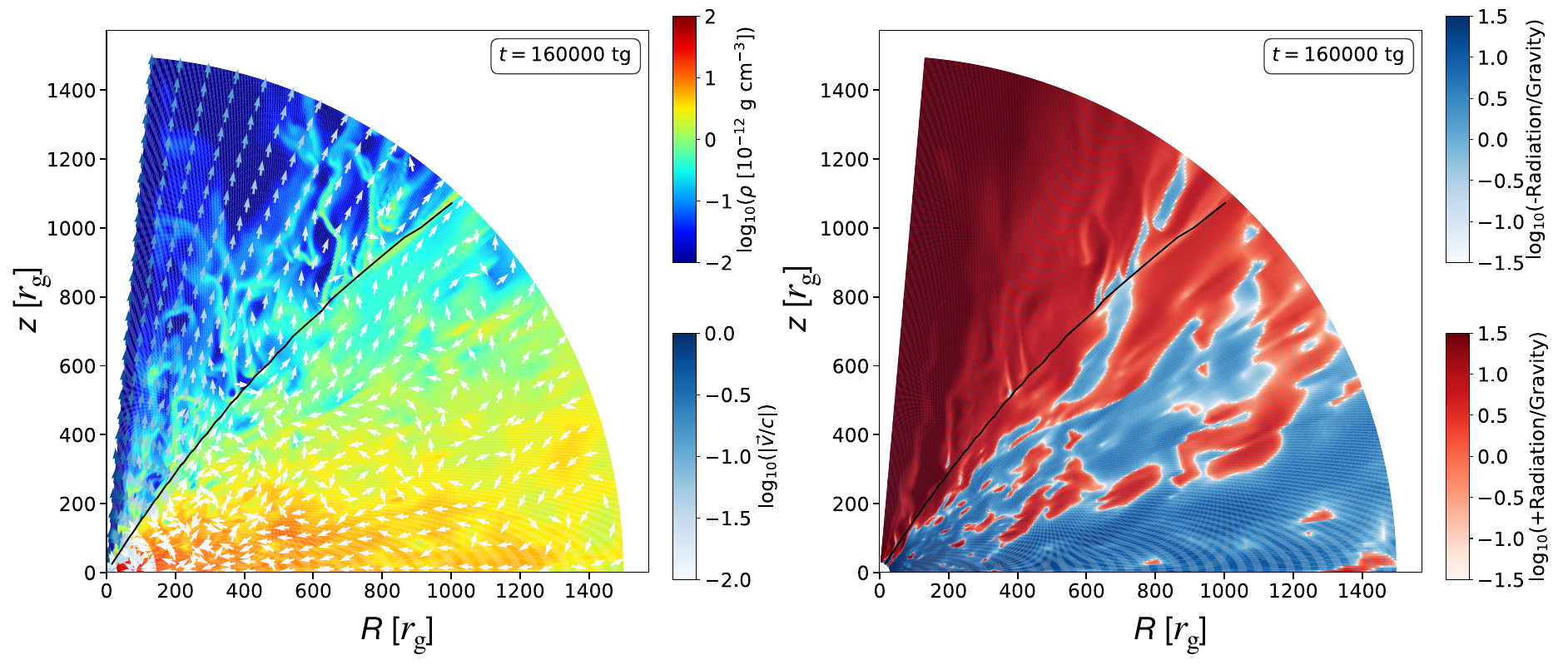}
 \end{center}
 \caption{Left: The same figure as Figure~\ref{fig_ICBC} but for time snapshot $t=1.6\times 10^5~r_{\rm g}/c$. Although highly dynamical, 
 the outflow structures are quasi-steady at this time. The clumpy outflows are clearly visible. Right: 2D distribution for force ratio 
 (radiation force/gravity). 
 The red regions indicate areas where outward radiation forces dominate, while the blue regions correspond to areas where inward gravity is predominant. The disk boundary can marginally capture the boundary between the inflow and outflow motion. In the sky region, radiation force are the main driver for the outflows.
 Alt text: Visualization of Simulations.}
 \label{fig_2Dplot}
\end{figure*}

The left panel of Figure~\ref{fig_2Dplot} presents the two-dimensional density distribution at $t=1.6\times 10^5~r_{\rm g}/c$ in the quasi-steady state. Above the orange and yellow regions representing the dense disk, low-density regions with outward velocities are spread out. These are the outflows.
The emergent outflows exhibit pronounced clumpy morphology and highly dynamic behavior as they propagate outward to larger spatial scales. In the right panel of Figure~\ref{fig_2Dplot}, we display the spatial distribution of the ratio between radiation and gravitational forces (considering only the radial component). Regions depicted in red correspond to outflowing material, while blue regions indicate inflowing matter. The disk boundary (black curve) effectively delineates the regions of inflow and outflow.
These results clearly demonstrate that radiation pressure serves as the dominant mechanism for launching and accelerating the outflows in our simulations.

\subsection{Clumpy Outflows: Observational Analogues}
\label{sec:3_2}

In our simulations, gas supplied from either the disk boundary or the inner boundary is accelerated by radiation and fragments, thereby generating clumps above the disk surface (see  Figure~\ref{fig_2Dplot}). These clumps are broadly consistent with the observations in terms of size, velocity, density, and spatial distribution \citep{Xrism+2025}. Here we note that, in this section, we define gas with ionization parameters
\begin{equation}
\xi = \frac{4\pi F_{r}}{\rho/\mu m_{\rm p}}
\label{eq:ionizationparametert}
\end{equation}
satisfying $3\lesssim \log(\xi/[{\rm erg~s^{-1}~cm}]) \lesssim 5$ as “clumps”.
This definition does not significantly contradict the XRISM observations of PDS 456, in which clumpy winds with $\log (\xi/[{\rm erg~s^{-1}~cm}]) \sim 5$ have been reported \hhBlue{(see Extended Data Table 1 in \cite{Xrism+2025} for detailed properties of clumpy outflows in PDS 456)}. \hhBlue{We find that slight difference in the definition of clumps won't change our main results. Nevertheless, we discuss the effects of different selection criteria for clumps using alternative ionization parameter choices in Section \ref{sec:ionizationcheck}.} \hhBlue{We also note that in simulation, radiative flux $F_{r}$ represents the total flux covering all wavelength for a black body spectrum. This approximation is supported by two evidences: (1) disk temperatures in our simulations are in range $10^5-10^6~{\rm K}$ (or higher, see temperature profiles in Figure 2 of \cite{Yoshioka+2024}), corresponding to radiation peak $\geq 1.8-18~{\rm Ryd}$; (2) spectrum energy distribution synthesis in \citet{Pacucci+Narayan2024} has pointed out that radiation from super-Eddington accreting disks peaks in range of $10^{15}-10^{18}~{\rm Hz}$, i.e., $\sim 0.3-300 ~{\rm Ryd}$, taking into account high-energy photons generated in outflows via inverse Compton scattering. }

\begin{figure*}[htbp]
 \begin{center}
  \includegraphics[width=115mm]{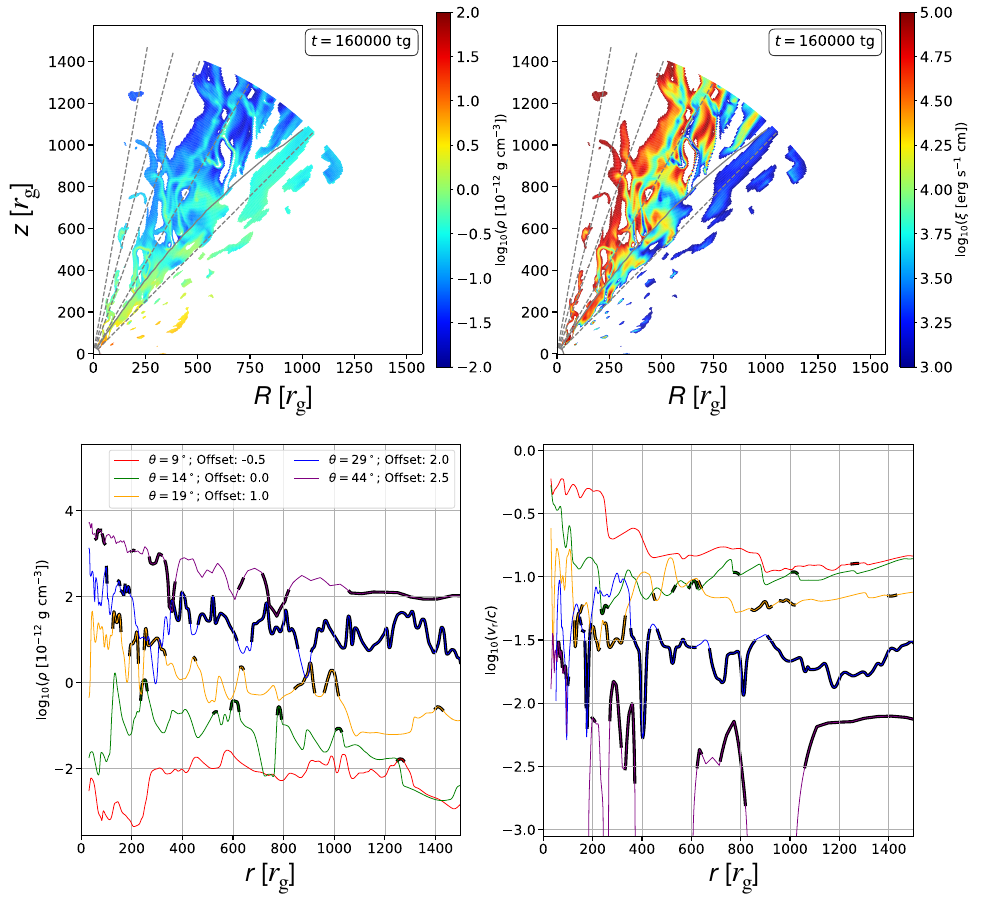}
 \end{center}
 \caption{\hhBlue{Top panels: 2D distribution of density and ionization parameters for clumpy outflows}. These clumps are selected such that their ionization parameters are within the observational range. Dashed gray curves show five directions at polar angles: $9^\circ$, $14^\circ$, $19^\circ$, $29^\circ$, and $44^\circ$,
 while solid gray curve indicates the disk boundary. \hhBlue{Bottom panels: radial profiles along the 5 polar angles for density (left) and outflowing velocity (right).} For better visualization, density profiles are offset by different values, as indicated in the legend. Along each radial profile, the selected clumpy outflow is highlighted with bold black curves.
 Alt text: Properties for clumpy outflows.}
 \label{fig_clumpyoutflow}
\end{figure*}

The \hhBlue{top panels of Figure~\ref{fig_clumpyoutflow} {
show} the 2D distribution of density and ionization parameters for {clumps selected based on the ionization parameter.}} 
\hhBlue{In practice, we construct the ionization parameters for all grid cells (e.g., the top-right panel in Figure~\ref{fig_clumpyoutflow}) and pick up cells that satisfy our clump selection criterion to build the density distribution.} As seen in these panels, clumps appear preferentially above the disk boundary, more specifically within the angular range $\theta \sim 10^\circ°-40^\circ$, and are distributed over a wide radial extent. 

To investigate the clump properties in more detail, we show in bottom panels of Figure~\ref{fig_clumpyoutflow} the density and radial (line-of-sight) velocity profiles along five polar angles ($\theta = 9^\circ, 14^\circ, 19^\circ, 29^\circ,$ and $44^\circ$). \hhBlue{Along each line of sight, density and outflowing radial velocity are plotted as profiles with thin solid curves. The ionization parameter is computed simultaneously along that line-of-sight, and regions that satisfy our clump-selection criterion are highlighted as thick black segments in the profiles.}
In the \hhBlue{bottom left} panel, offsets are added to the five profiles for better visualization. A single clump is clearly identified along $\theta = 9^\circ$, while multiple clumps are clearly identified along $\theta = 14^\circ$ and $19^\circ$. At $\theta = 29^\circ$, the ionization parameter satisfies the criterion over a more extended radial region compared to other angles. This means that individual clumps become larger in size, but the number of clumps does not increase significantly. The clumps appear as local density maxima in the radial profiles, with typical sizes of a few to several tens of $r_{\rm g}$. The \hhBlue{bottom} right panel shows that these clumps achieve outflow velocities in the range of $0.05$--$0.2~c$, which are slightly lower or comparable to the observational values \hhBlue{(see Extended Data Table 1 in \cite{Xrism+2025} for velocities of clumpy outflows in PDS 456)}. Note that the results along $\theta = 44^\circ$ are strongly affected by the disk boundary condition beneath the disk surface, and thus may not represent the physical outflow properties, but are shown for reference. \hhBlue{Combining the right panels, we notice a positive relation between wind velocity and ionization parameters by visual inspection: high velocity winds occurs at smaller polar angles where ionization parameters are tend to be higher (redder in the top {right} panel) as noted in \cite{Xu+2025} for PDS 456. However, we refrain from further interpretation due to simplified radiation transfer treatment in current simulations. }

\begin{figure}[htbp]
 \begin{center}
  \includegraphics[width=78mm]{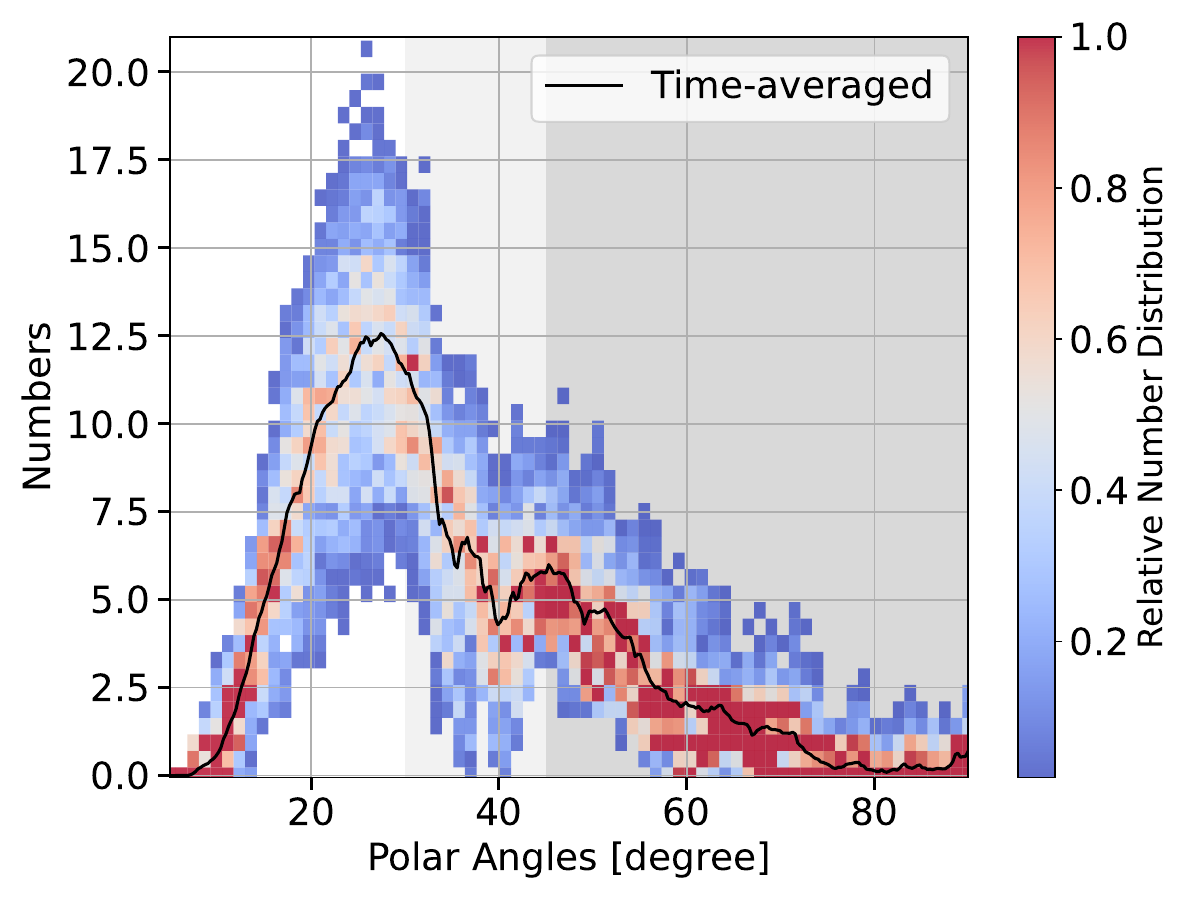}
 \end{center}
 \caption{The angular number distribution for clumpy outflows in simulation. The color-coded regions are integrated distribution for all data snapshots ranging from $t=0.9\times 10^5~r_{\rm g}/c$ to $t=1.7\times 10^5~r_{\rm g}/c$, with time step $\delta t=200 ~r_{\rm g}/c$, while the solid black curve is the time-averaged number distribution. Note that for polar angles $30^\circ \lesssim \theta \lesssim 45^\circ$, the line-of-sights penetrate through the disk region (light-gray shaded region), and for $\theta \gtrsim 45^\circ$, line-of-sights are entirely embedded in disk region (gray shaded region).
 Alt text: Angular number distribution.}
 \label{fig_angular}
\end{figure}

Figure~\ref{fig_angular} presents the angular distribution of clump numbers. Here, the clump number is defined as the number of regions along a given line of sight (at constant $\theta$) where the ionization parameter lies within the criterion range. For example, from the density profile at $\theta = 14^\circ$ in the middle panel of Figure~~\ref{fig_clumpyoutflow}, the number of clumps is found to be five. We calculate the clump number for each polar angle at every time snapshot during $0.9$--$1.7 \times 10^{5}\, ~r_{\rm g}/c$ with a temporal resolution of $\delta t = 200\, ~r_{\rm g}/c$, and display the resulting relative number distributions as color maps, while the black curve shows the time-averaged distribution. It should be noted that a single clump may be counted at multiple polar angles if it extends over them. Thus, Figure~\ref{fig_angular} represents the ``observed'' angular clump number distribution rather than the intrinsic one. Also, for $\theta \geq 30^\circ$, the lines of sight pass through the lower part of the disk boundary, making the results in this region less reliable.


From this figure, a prominent peak is seen at $\theta \sim 20^\circ$--$30^\circ$, while the number of clumps decreases toward both smaller angles (toward the rotation axis) and larger angles (toward the equatorial plane). A secondary peak is also seen inside the disk region. The average number of clumps along a single line of sight is about seven, which is broadly consistent with the observation (five).

Since at least one clump exists in the range of $\theta = 10^\circ$–$40^\circ$ above the disk, the covering factor of clumps is 
\begin{equation}
\frac{\Omega_{\rm clumps}}{4\pi} = \int_{\theta_1}^{\theta_2} 2\cdot\sin(\theta){\rm d}\theta \sim 0.44,
\end{equation}
\hhBlue{with $\theta_1=10^\circ$ and $\theta_2=40^\circ$}.
However, if the gas also fragments in the azimuthal direction and larger gaps are formed, this value could be reduced further.

\begin{figure*}[htbp]
 \begin{center}
  \includegraphics[width=170mm]{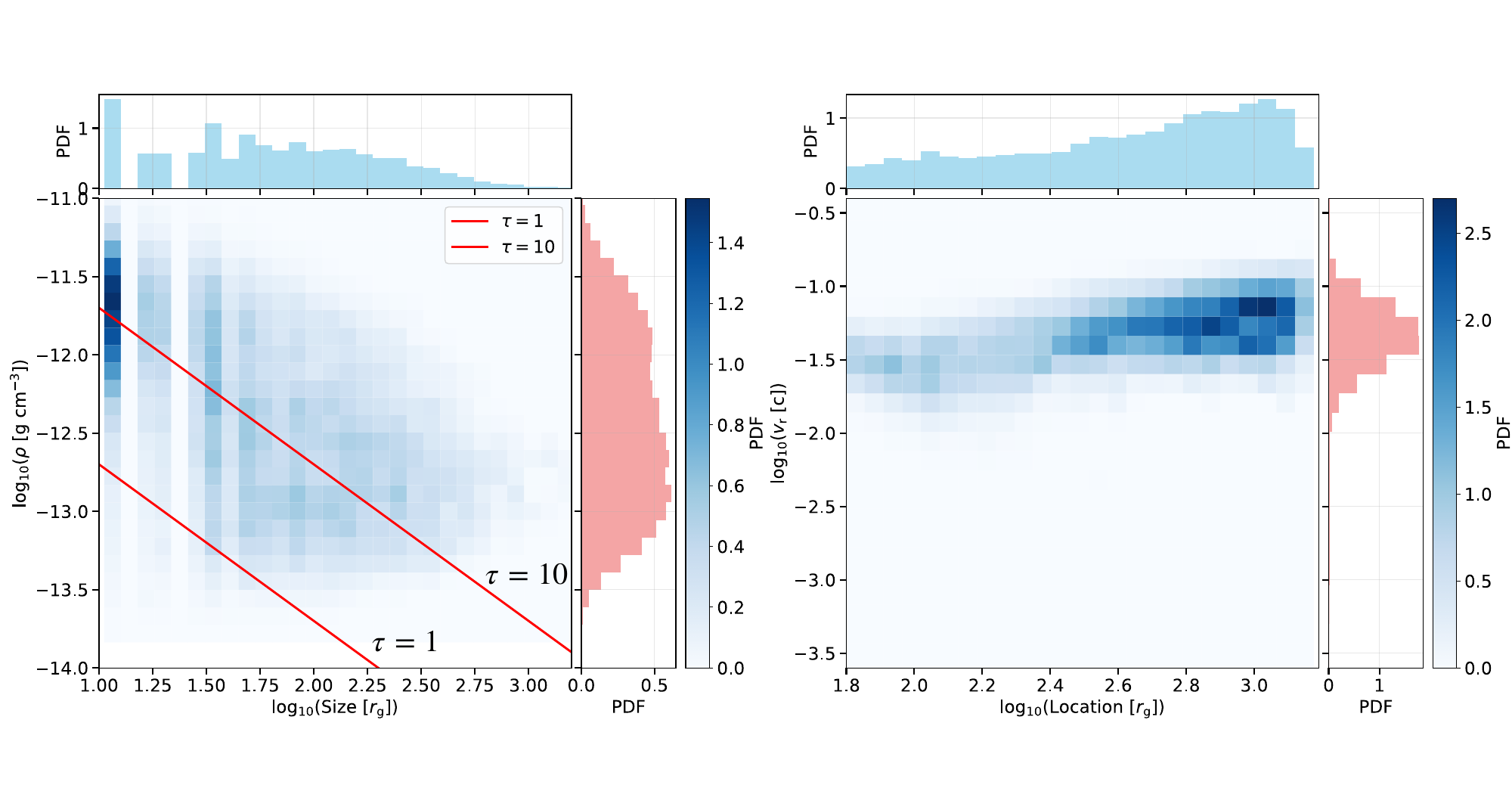}
 \end{center}
 \caption{Statistical distribution of properties for clumpy outflows. Left: Distribution in the clump density and clump size plane. The probability distributions for density and size are shown as side plots in the right and upper side of this panel, respectively. The solid red curves indicate the relation where $\tau = \rho \kappa_{\rm es}\Delta r=1$ and $10$.
 Right: The same distribution as left panel but for clump radial velocity and clump location plane. 
 Alt text: Statistical properties for clumpy outflows.}
 \label{fig_statistics}
\end{figure*}

\subsection{Statistics of Clumpy Outflows}\label{sec:3_3}

In this subsection, we adopt a statistical approach to characterize the ensemble properties of these clumpy outflows, providing insight into their collective evolution and dynamics. \hhBlue{Following the similar procedures as done in Figure~\ref{fig_clumpyoutflow}, clumps are identified and picked up along each polar angles (with { 
an interval of $\delta \theta=\pi/2N_{\theta}$}) when ionization parameters met our criterion for each time snapshot. After clump identification, we systematically calculate their key physical properties, including their characteristic {radial} sizes, mean velocities, average densities, and radial distances from the central radiation source.}
This analysis encompasses clumps detected across all polar angles and throughout the temporal domain spanning $0.9-1.7\times 10^5~r_{\rm g}/c$, sampled at intervals of $\delta t=200~r_{\rm g}/c$. The resulting statistical distributions of these properties are presented in Figure~\ref{fig_statistics}.

The left panel of Figure~\ref{fig_statistics} presents the joint probability distribution of clump properties in the density-size plane, with the corresponding marginalized probability distributions for density and size displayed in the right and upper side panels, respectively. The red solid curves overlaid on the main panel represent contours of constant optical depth, corresponding to $\tau = \rho \kappa_{\rm sca}\Delta r = 1$ and $\tau = 10$, where $\kappa_{\rm sca}$ denotes the electron scattering opacity and $\Delta r$ represents the clump size. 

This visualization reveals that the majority of identified clumps are concentrated in the the region with optical depths in the range of $\tau \sim 1-10$. The density distribution (right side panel) exhibits a concentration of probability distribution at approximately $10^{-12}$ and $10^{-13}~{\rm g~cm^{-3}}$. Regarding the size distribution (upper panel), the clumps predominantly exhibit characteristic dimensions spanning $10-100~r_{\rm g}$, with a smaller subset extending to larger sizes of approximately $300~r_{\rm g}$. \hhBlue{As a showcase in Figure~\ref{fig_clumpyoutflow}, several clumps with size of order $\sim 10~r_{\rm g}$ are found to emerge around $\sim 200-1400~r_{\rm g}$ (e.g., along polar angles $9^\circ,~14^\circ,~19^\circ$). It is also found that clumps found at large polar angles are tend to be larger in size (e.g., along $\theta\sim 29^\circ,~44^\circ$). It may have traced clumps before their fragmentation.}

The right panel of Figure~\ref{fig_statistics} illustrates the joint distribution of clump properties in the velocity-radius plane, with the corresponding marginalized probability distributions for outflow velocity and radial distance from the central radiation source displayed in the right and upper side panels, respectively.

Examining first the velocity distribution (right side panel), we observe an approximately log-normal distribution centered at $10^{-(1.3-1.4)}~c \approx 0.04-0.05~c$. This distribution indicates that the majority of identified clumps propagate at significant velocities, with more than half exceeding $0.05~c$. 
A small fraction of the clumps reach velocities as high as $0.2~c$.
The radial distance distribution (upper panel) exhibits a monotonic increase toward larger scales (extending to $\sim 1000~r_{\rm g}$), a trend that likely reflects the geometric effect of increasing volume with increasing radius.

In the velocity-radius plane (main panel), we identify a positive correlation between outflow velocity and radial distance, with a notable concentration of clumps in the upper-right region of the parameter space. This correlation suggests that clumps continue to experience significant outward acceleration due to radiation pressure as they propagate to larger radii. A substantial fraction of the clumps reside at distances of $200-1000~r_{\rm g}$ from the central source, where they remain subject to radiative acceleration.

It is noted again that the statistical distributions presented here characterize the ``observed" properties of ionization-parameter-selected clumps along all lines-of-sight, rather than representing the intrinsic statistical properties of physically distinct clumps. This distinction arises because individual clumps that span multiple polar angles are counted separately along each line-of-sight they intersect, potentially leading to multiple representations of the same physical structure in our statistical analysis. Despite this methodological consideration, 
the ``observed" clumpy outflows in our simulations typically exhibit characteristic sizes of $10-100~r_{\rm g}$, propagate with velocities in the range of $0.05-0.2~c$, and extend radially from approximately $100~r_{\rm g}$ to $1000~r_{\rm g}$ from the central source. 
The clump sizes obtained in the simulations are slightly larger, and the velocities are somewhat lower, but they are broadly consistent with the XRISM observations \hhBlue{(see Extended Data Table 1 in \cite{Xrism+2025})}, which report sizes of $2-16 ~r_{\rm g}$, velocities of $0.2-0.3~c$, and radial distances of $200-600 ~r_{\rm g}$.
This consistency suggests that radiation-driven, clumpy outflows generated in super-Eddington accretion systems represent a viable physical mechanism for producing the multi-velocity absorption features detected in recent high-resolution X-ray spectroscopic observations \citep{Xrism+2025,Xu+2025}.

The shift of the clump sizes toward larger values may be due to the insufficient numerical resolution. Higher-resolution simulations will be necessary in future work. The discrepancy in velocities might be resolved by adjusting the mass accretion rate. It has been point out that \citep{Yoshioka+2022,Yoshioka+2024}, although the outflow density decreases as the accretion rate decreases, the velocity tends to increase (see also \cite{Hu+2022}). In addition, if absorption lines are produced by gas with higher ionization parameters than those adopted in this study, the clumps may have higher velocities, because gas closer to the rotational axis tends to have higher ionization parameters and higher velocities. A detailed comparison with observations will be addressed in future work.

\section{Discussions}\label{sec:discussion}

\subsection{Prerequisites for Reproducing Clumps}

\begin{figure*}[htbp]
 \begin{center}
  \includegraphics[width=170mm]{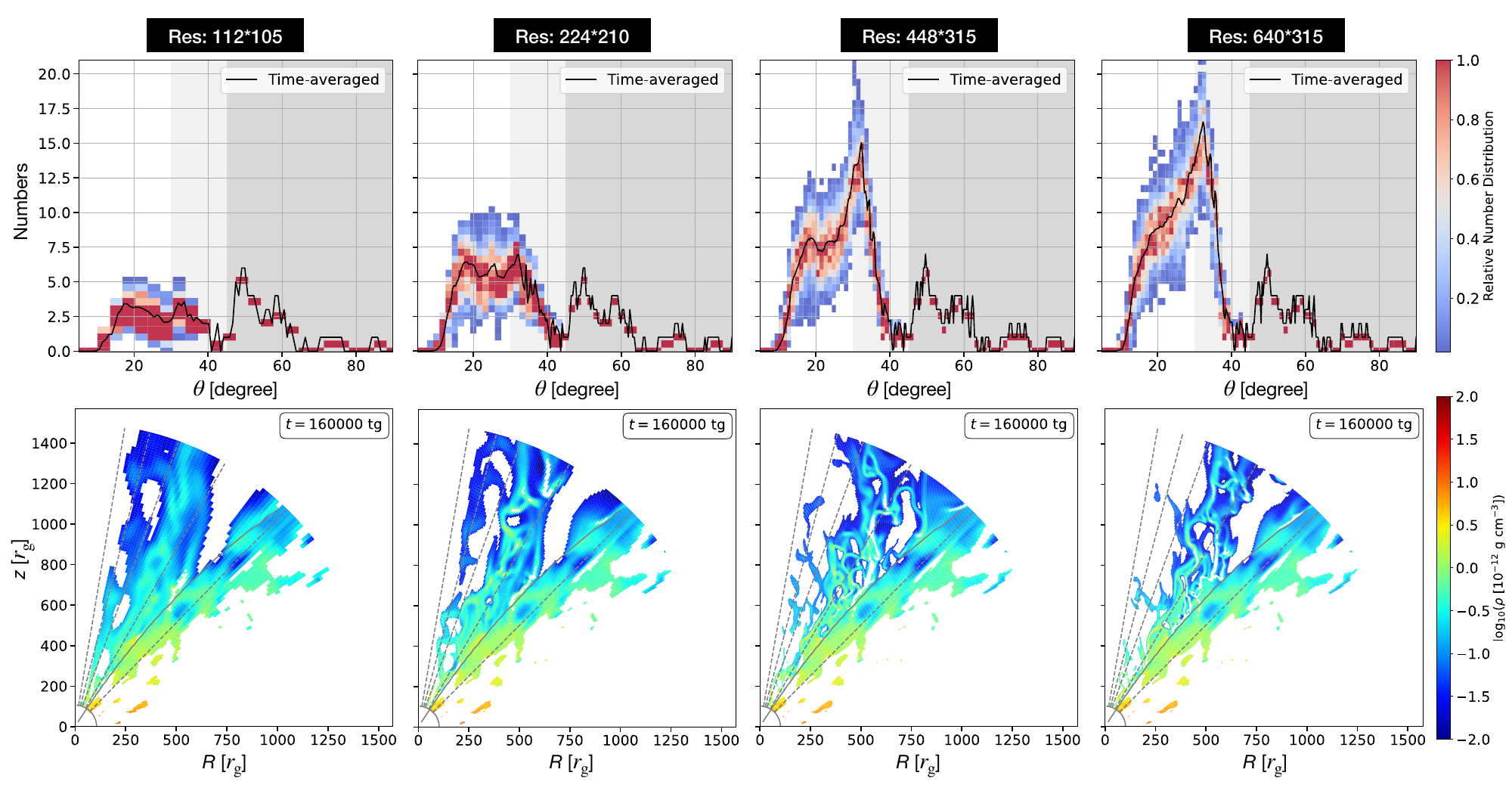}
 \end{center}
 \caption{Top panels: Angular number distribution for simulations with various grid resolutions. From the left to right, the resolutions are: ($N_{r}$, $N_{\theta}$) = $(112,~105)$, $(224,~210)$, $(448,~315)$, $(640,~315)$. Bottom panels: ionization-parameter-selected outflow structures. In these simulations, inner boundary conditions and disk boundary conditions are fixed (see time-varying conditions for simulation case in Sec.~\ref{sec:CO}), to avoid effects due to time-varying conditions.
 Alt text: Resolution Check.}
 \label{Fig:ResolutionCheck}
\end{figure*}

We here point out two important facts regarding the generation of clumpy winds:
(1) Clumpy winds can emerge even without time-dependent boundary conditions; (2) If the numerical resolution of the simulations is insufficient, the clumpy winds either fail to appear or cannot be accurately reproduced.

In the fiducial simulations discussed in Section 3, the physical quantities at the inner and disk boundaries vary with time, whereas in the test simulations shown in Figure~\ref{Fig:ResolutionCheck}, these quantities were fixed at their initial values without temporal variation. The upper panels display the time-averaged clump number distributions. The lower panels show the corresponding two-dimensional density distributions of ionization-parameter-selected clumps from a representative snapshot ($t=1.6\times 10^5~t_{\rm g}$) for each simulation. 
The grid configurations used are ($N_{r}$, $N_{\theta}$) = $(112,~105)$, $(224,~210)$, $(448,~315)$, and $(640,~315)$, where $N_r$ and $N_\theta$ represent the number of radial and polar grid cells, respectively.
In the rightmost panel, which has the same resolution as the fiducial simulation, it is found that clumpy winds are formed.
Furthermore, a comparison with the results of the fiducial simulation shown in Figure~\ref{fig_angular} indicates that
the statistical properties of the clump populations are quite similar throughout most of the computational domain. The overall consistency  demonstrates that the fundamental physical mechanisms driving clump formation operate effectively even in the absence of temporal boundary variations. This finding indicates the emergence of clumpy outflows is robust to these boundary details. 

\begin{figure*}[htbp]
 \begin{center}
  \includegraphics[width=110mm]{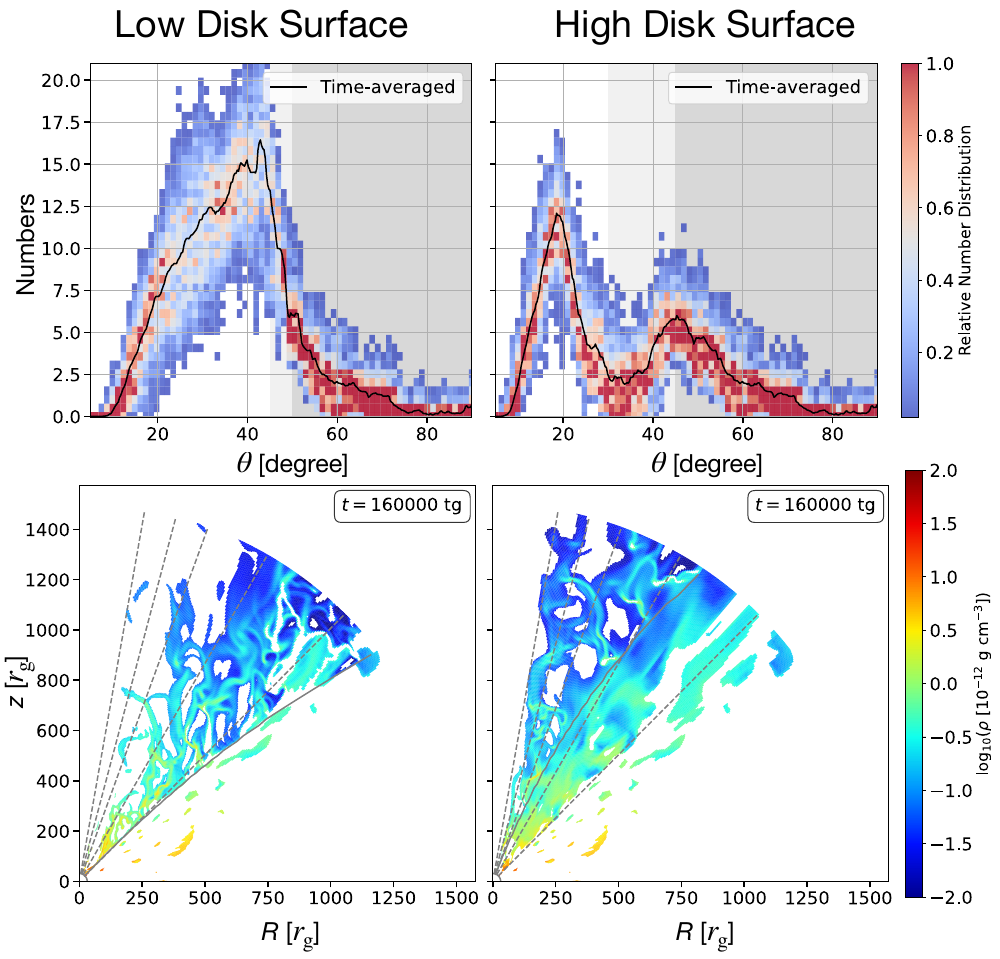}
 \end{center}
\caption{The same figure as Figure~\ref{Fig:ResolutionCheck} but for simulation cases with different disk boundary.
Disk surfaces are indicated in solid gray curves in the bottom panels. These disk surfaces are artificially chosen such that they are slightly below, and above the actual disk surface (where gravity is balanced by radiation force, see \cite{Yoshioka+2024}). The shaded regions are similar to that in Figure~\ref{fig_angular}. 
Alt text: Results for different disk boundary.}
 \label{Fig:DiskSurfaceCheck}
\end{figure*}

In the lowest-resolution simulation, ($N_{r}$, $N_{\theta}$) = $(112,~105)$, the structures identified as clumps appear more like continuous outflow streams (left panel of Figure~\ref{Fig:ResolutionCheck}). In contrast, in the higher-resolution simulations, outflows with more complex substructures clearly emerge. This result implies that clumpy outflows cannot be reproduced unless the resolution is sufficiently high. In addition, the number of resolved clumps increases with resolution (see the left three panels) but converges at higher resolutions (compare the two rightmost panels). \hhBlue{We do not find significant different in clump properties as long as clumps are resolved with sufficient resolution.}
These findings demonstrate that the grid resolution of the fiducial simulation is sufficient to reproduce the statistical properties of clumpy outflows. The limited reports of clumpy structures in previous radiation-hydrodynamic simulations \citep{Takeuchi+2013,Kobayashi+2018} may have been due to ``insufficient resolution”.
\hhBlue{We also note that since spherical coordinate system is adopted in our simulation, resolution at $\theta$-direction becomes coarse at large radii, e.g., $\delta\cdot\theta r_{\rm out}\gg \delta \theta \cdot r_{\rm in}$. To this concern, our discussions are limited to the radial extent of clumps rather than their angular sizes (nor their azimuthal distribution). }

Although identifying the precise physical origin of clump formation is beyond the scope of this paper, several mechanisms have been suggested for clumpy outflows in radiation-dominated environments, including thermal instability \citep{Field1965}, photon-bubble instability \citep{Arons1992}, radiation-induced instability, and Rayleigh-Taylor instability \citep{Shaviv2001,Takeuchi+2013,Kobayashi+2018}. A detailed investigation of the clump formation mechanism is left as an important future work. 

\begin{figure*}[htbp]
 \begin{center}
  \includegraphics[width=110mm]{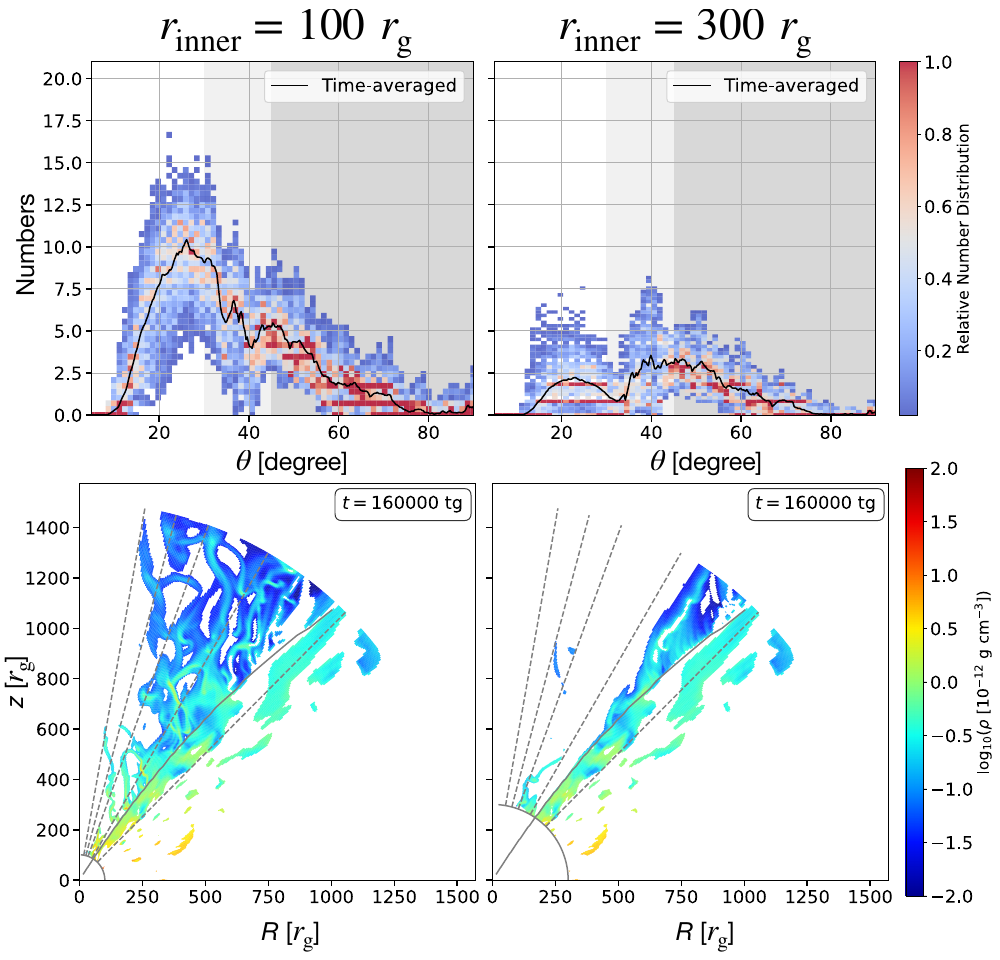}
 \end{center}
\caption{The same figure as Figure~\ref{Fig:DiskSurfaceCheck} but for simulation cases with different inner boundaries. The left and right panels are results for simulation with $r_{\rm inner} = 100~r_{\rm g}$, and $r_{\rm inner} = 300~r_{\rm g}$, respectively. We find that there is a minor difference between left and the fiducial case (see Section~\ref{sec:CO}), while the right panel differ significantly from the fiducial case. There is almost no clump detected in the right panel.
Alt text: Results for different inner radii.}
 \label{Fig:InnerRadiusCheck}
\end{figure*}

\subsection{Dependence on Disk Boundary Conditions}

As outlined in Section~\ref{sec:setup}, our 
simulations focus specifically on the outflow dynamics rather than accretion disk evolution. Therefore, by introducing a disk boundary, we effectively exclude the disk region from the computational domain. 
This boundary is artificial and we discuss this effect in this subsection.

In practice, in addition to the fiducial model, we performed simulations with two different disk boundaries: one with a lower boundary (“Low Disk Surface”) and the other with a higher boundary (“High Disk Surface”).
In Figure~\ref{Fig:DiskSurfaceCheck} we present the angular number distribution of clumps for these two simulations (top panels) alongside corresponding 2D density distributions for ionization-parameter-selected clumps (bottom panels). As shown in Figures \ref{fig_clumpyoutflow} (\hhBlue{top left panel}) and \ref{Fig:DiskSurfaceCheck}, clumpy winds appear above the disk boundary (in the sky region) for all three simulations. As seen from Figures \ref{fig_angular} and \ref{Fig:DiskSurfaceCheck}, in all three cases, the number distribution increases with distance from the rotation axis and then decreases as it approaches the disk boundary. Although the peak positions differ because of the different locations of the disk boundaries, 
the overall trend remains the same in all models.
It indicates the clump properties in the sky regions are robust and reliable, though the location of the disk boundary does have important effects on clump properties. 
However, we refrain from drawing more definitive conclusions regarding disk boundary conditions because disk scale heights are intrinsically sensitive to BH accretion rates, while our boundary implementations remain simplified and artificial. 
To investigate the effects of different accretion rates and disk configurations, the disk boundary must be removed so that the entire disk region is included in the computational domain. In such simulations, magnetic fields should also be incorporated \hhBlue{(e.g., \cite{Fukumura+2015}).}

\begin{figure*}[htbp]
 \begin{center}
  \includegraphics[width=110mm]{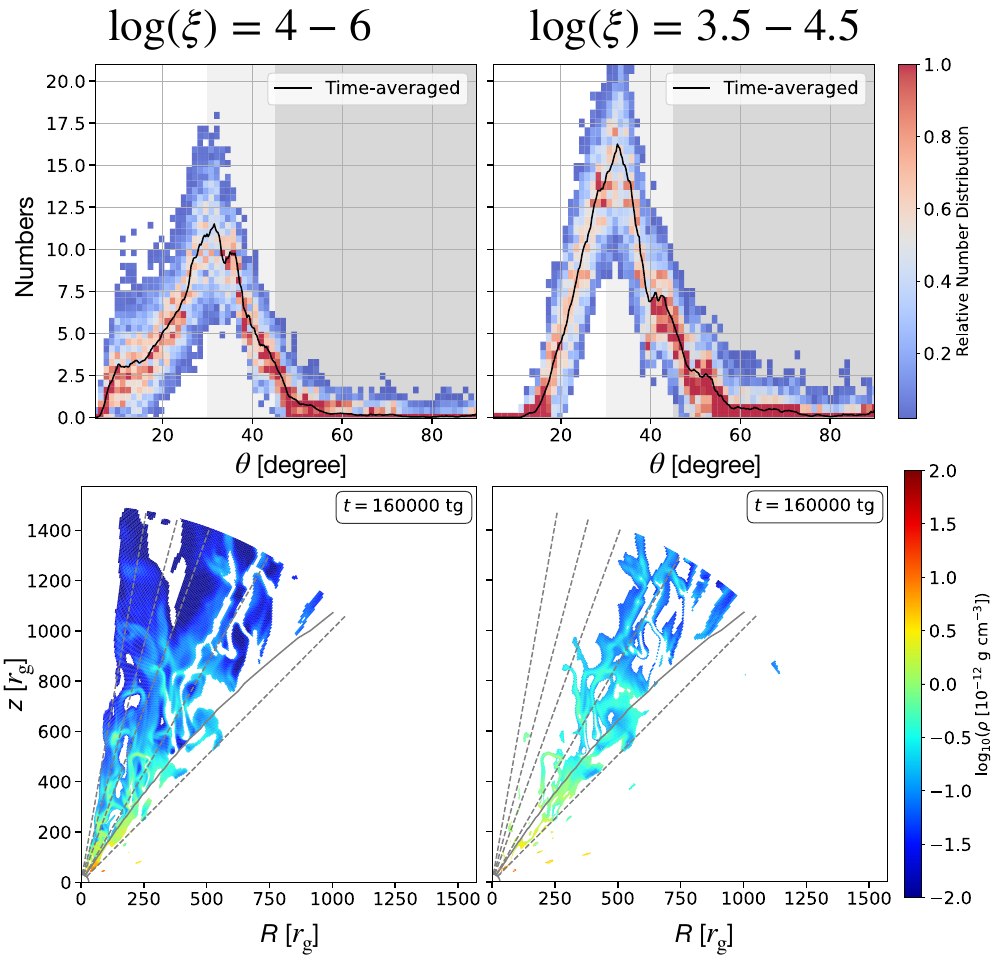}
 \end{center}
\caption{The angular number distribution and 2D density distribution for fiducial simulation with different clump selection criteria. 
The left panel shows the results for the ionization parameter range of $\log(\xi /{\rm erg~s^{-1}~cm}) = 4$–$6$, and the right panel shows those for $3.5$–$4.5$. In both cases, high-density regions are identified as clumps, while in the left case with a higher ionization parameter selection criterion, many low-density regions are also miscategorized as clumps.
Alt text: Results for different selection criteria.}
 \label{Fig:IonizationDegreeCheck}
\end{figure*}

\subsection{Dependence on Inner Boundaries}

Despite the aforementioned prerequisites on numerical setups, our investigation reveals that varying the inner boundary radius produces substantially different results in the simulated outflow properties. 
Figure~\ref{Fig:InnerRadiusCheck} shows the results for two cases with different inner radii from our fiducial case: $100~r_{\rm g}$ (Left panel), and $300~r_{\rm g}$ (Right panel). Note that changing the location of the inner boundary also alters the inner edge position of the disk boundary. For each simulation, we implement inner and disk boundary conditions derived from the physical quantities at the corresponding locations in RHD simulations from \citet{Yoshioka+2024}.

Comparing the fiducial case (see Figures \ref{fig_clumpyoutflow} and \ref{fig_angular} in Section~\ref{sec:CO}) with results presented in the left panel of Figure~\ref{Fig:InnerRadiusCheck}, both cases clearly exhibit clumpy structures above the disk boundary. In addition, we observe only minor differences in the angular number distributions. The increase in the inner radius only slightly lower the peak of the number distribution due to the exclusion of the $30-100~r_{\rm g}$ region in the computational domain. In stark contrast, the right panel ($r_{\rm inner}=300~r_{\rm g}$) exhibits a dramatically different distribution with almost no detectable clumpy structures. It is also evident from the density distribution that the clumpy structures have not appeared (see the lower-right panel of Figure~\ref{Fig:InnerRadiusCheck}). 

These results implicitly indicate that the treatment of the region within $r\lesssim 300~r_{\rm g}$ is crucial for clump formation. In this region, outflows moving radially from the vicinity of the BH interact with gas launched toward the rotation axis from the disk boundary, and this interaction appears to produce clumpy winds.
The simulations with $r_{\rm inner}=30~r_{\rm g}$ and $r_{\rm inner}=100~r_{\rm g}$ accurately resolve this process, thereby leading to clump formation. In contrast, although a similar interaction is expected to occur in the $r\lesssim 300~r_{\rm g}$ region of the simulation by \citet{Yoshioka+2024}, its lower spatial resolution likely prevents this process from being properly captured. Consequently, when the results of that simulation are adopted as the boundary condition, clumps do not appear in the $r_{\rm inner}=300~r_{\rm g}$ run.
In summary, resolving the interactions within $r\lesssim 300~r_{\rm g}$ at sufficiently high spatial resolution is essential for reproducing clump formation.

\hhBlue{However, we caution the degeneracy between the changes of inner boundaries and radiation at inner boundary. Our simulation setups do not allow free modifications to radiation field, for the sake of physical consistency of initial and boundary conditions. Thus, it is hard to tell if the aforementioned stark differences are due to changes in radiation or other physical processes at small scales ($r\lesssim 300~r_{\rm g}$). A comprehensive investigation of this degeneracy will be addressed in our forthcoming research. }

\subsection{Ionization Parameter Selection Criterion}
\label{sec:ionizationcheck}

In our fiducial case, we adopted the ionization parameter range of $3 \lesssim \log(\xi / {\rm erg~s^{-1}~cm}) \lesssim 5$. In order to evaluate the sensitivity of the results to this selection criterion, we perform additional analyses of the fiducial simulation applying two different ionization parameter ranges: a higher range of $4 \leq \log(\xi) \leq 6$ and a narrower range of $3.5 \leq \log(\xi) \leq 4.5$. Figure~\ref{Fig:IonizationDegreeCheck} presents the angular number distribution (top panels) and 2D density distribution (bottom panels) for clumps selected using these different ionization parameter criteria.

Comparing the case with the selection criterion of $4 \lesssim \log(\xi) \lesssim 6$ (lower-left panel) with the fiducial case (Figure \ref{fig_clumpyoutflow}), we find that the selected regions become more extended, which is due to the inclusion of low-density, highly ionized regions. However, the angular number distribution (see Figure 4 and the upper-left panel) shows that it does not change significantly.
When adopting the narrower selection range ($3.5 \leq \log(\xi) \leq 4.5$, the right panel), primarily high-density regions are extracted. As a result, the clump sizes become visibly smaller, although the angular number distribution of clumps remains similar to that in the fiducial case.

Our selection criterion, which is consistent with observations such as $\log(\xi/[{\rm erg~s^{-1}~cm}]) \sim 5$ reported by the XRISM Collaboration (2025), successfully separates clumps from diffuse outflows. However, a detailed comparison between observations and simulations requires performing radiation transfer calculations to synthesize spectra (e.g., \cite{Sim+2010}; \cite{Mizumoto+2021}; \hhBlue{\cite{Hagino+2015}}).

\section{Conclusions}\label{sec:conclusion}

In this paper, we performed radiation hydrodynamic simulations to investigate the formation mechanisms and evolution of clumpy outflows around super-Eddington accreting supermassive black holes. Using the results of \citep{Yoshioka+2024} as the initial and boundary conditions, we carried out high-resolution simulations focusing on the radiation-driven outflow region above the accretion disk.

The obtained clumpy outflows are widely distributed both in the radial and polar directions, with a relatively large covering factor of $\sim 0.2$. The typical clump sizes range from $10$ to $100~r_{\rm g}$, their densities are on the order of $10^{-12}$–$10^{-13}~{\rm g~cm^{-3}}$, and their optical depths for electron scattering are approximately $1$–$10$. Their velocities are in the range of $0.05$–$0.2~c$.
Along a line of sight covering $100-1000 ~r_{\rm g}$, about five clumps are typically found. These results are broadly consistent with the observational findings for PDS 456 \citep{Xrism+2025}, although the simulated velocities are slightly lower.
This velocity discrepancy could potentially be resolved by adjusting the mass accretion rate and disk luminosity, which regulate the radiation-driven acceleration. It is also possible that additional physical effects, such as magnetic fields, BH accretion rate, and absorption line ionization, contribute to the acceleration. Further investigation of these mechanisms is left for future work.

To reproduce the clumpy outflows, sufficiently high grid resolution and non-uniform boundary conditions that introduce perturbations are required. 
The interaction between the outflows moving in the radial direction and the gas launched from the disk surface, which occurs within approximately $300~r_{\rm g}$, is inferred to play an important role in the formation of the clumpy structure.
While this paper focuses on reproducing observed features rather than determining formation mechanisms, a forthcoming companion paper will comprehensively address the physical processes driving clump formation.

In this study, we adopted an observation-based criterion for identifying clump structures (ionization parameter of $10^{3-5}$ erg s$^{-1}$ cm). However, a more accurate comparison between theory and observations requires performing radiation transfer calculations to obtain the emergent spectra. In addition, fully three-dimensional simulations will be necessary to properly capture the clump morphology and the propagation of radiation.

\begin{ack}
We are grateful to anonymous referees for their valuable and constructive comments, which helped us improve the manuscript.
The numerical calculations and simulations were performed partially on HPE Cray XD2000 at the Center for Computational Astrophysics (CfCA), National Astronomical Observatory of Japan and partially on Yukawa-21 at Yukawa Institute for Theoretical Physics (YITP) in Kyoto University.
\end{ack}

\section*{Funding}
This research was supported by the Japan Society for the Promotion of Science (JSPS) through KAKENHI Grant Numbers 24KF0130 (HH, KO), JP21H04488(KO), 24K00678(KO, HRT),  25K01045 (KO), 24K00672 (HRT) and 23K03445 (YA). This work was also supported by MEXT as “Program for Promoting Researches on the Supercomputer Fugaku” (Structure and Evolution of the Universe Unraveled by Fusion of Simulation and AI; Grant Number JPMXP1020240219; YA, HRT, KO), 
by Joint Institute for Computational Fundamental Science (JICFuS, KO), and
(in part) by the Multidisciplinary Cooperative Research Program in CCS, University of Tsukuba. 
This work was also supported in part by JST SPRING Grant No. JPMJSP2110 (SY).

\section*{Data availability} 
 The data underlying this article are available under reasonable request upon to the corresponding author.







\end{document}